\documentclass[12pt]{article}

\pdfoutput=1
\usepackage[margin=1in]{geometry}
\usepackage{amsmath,amssymb,mathrsfs,amsthm,graphicx,verbatim}
\usepackage{subfigure,float}
\usepackage{color}
\usepackage[sort]{natbib}
\usepackage{tikz}
\usepackage{booktabs}
\usepackage{multirow}
\usepackage{hhline}
\usepackage{varioref}
\usepackage{caption}
\usepackage{fullpage}
\usepackage{lineno}
\usepackage{placeins}
\usepackage{setspace}

\newcommand{\mbf}[1]{\mathbf{#1}}
\newcommand{\mbs}[1]{\boldsymbol{#1}}
\newcommand{\mcal}[1]{\mathcal{#1}}
\newcommand{\mbb}[1]{\mathbb{#1}}
\newcommand{\N}{\mbb{N}_0}

\newcommand{\by}{\mbf{y}}

\newcommand{\ba}{\mbf{a}}
\newcommand{\bb}{\mbf{b}}
\newcommand{\bx}{\mbf{x}}

\newcommand{\bX}{\mbf{X}}

\newcommand{\bbeta}{\mbs{\beta}}
\newcommand{\balpha}{{\mbs{\alpha}}}
\newcommand{\bepsilon}{{\mbs{\epsilon}}}
\newcommand{\bgamma}{{\mbs{\gamma}}}
\newcommand{\bpi}{{\mbs{\pi}}}
\newcommand{\bI}{\mbf{I}}
\newcommand{\bH}{\mbf{H}}

\newcommand{\1}{\mbs{1}}

\renewcommand{\exp}[1]{\text{exp}\left[#1\right]}
\newcommand{\M}{{M}}

\newcommand{\p}{{p}}
\newcommand{\MB}{{M_B}}
\newcommand{\MF}{{M_F}}
\newcommand{\MT}{{M_T}}
\newcommand{\cM}{\mathcal{M}}
\newcommand{\graph}{\Gamma}
\newcommand{\kset}{\Upsilon}
\newcommand{\order}{order}
\newcommand{\parents}{\mcal{P}}
\newcommand{\children}{\mcal{C}}
\newcommand{\extreme}{\mcal{E}}

\newcommand{\lrb}[1]{\left\{#1\right\}}

\newtheorem{Theorem}{Theorem}

\usetikzlibrary{shapes,shadows,arrows}
\tikzstyle{block1}=[draw,loosely dotted,circle,%
text width=2.4em,minimum height=3em, text centered, node distance=4em]
\tikzstyle{block2}=[draw,densely dashed,circle,%
text width=2.4em,minimum height=3em, text centered, node distance=4em]
\tikzstyle{block3}=[draw,circle,%
text width=2.4em,minimum height=3em, text centered, node distance=4em]
\tikzstyle{block4}=[draw,circle,fill=orange!80,text width=2.4em,minimum height=5mm, text centered, node distance=4em]
\tikzstyle{block5}=[draw,circle,fill=orange!0,text width=2.8em,minimum height=5mm, text centered, node distance=4em]
\tikzstyle{block6}=[draw,circle,fill=orange!0,text width=2.4em,minimum height=5mm, text centered, node distance=4em]
\tikzstyle{blocktext}=[fill=white,text width=1.5em,minimum height=5mm, text centered, node distance=3.5em]
\tikzstyle{blocktext2}=[fill=white,text width=3em,minimum height=10mm, text centered, node distance=3.5em]
\tikzstyle{line}=[draw,-latex']

\usepackage{bibunits}

\title{\vspace{-1in}Bayesian Variable Selection on Model Spaces Constrained by Heredity Conditions\\
{\large {\bf Short Title:} Variable Selection with Heredity Conditions}
}
\author{Daniel Taylor-Rodriguez
\and 
Andrew Womack 
\and
Nikolay Bliznyuk
}

\date{January 31, 2015}

\begin{document}
\defaultbibliography{BibLowerOrder}
 \defaultbibliographystyle{apalike}
\numberwithin{table}{section}
\numberwithin{figure}{section}

\maketitle

\setcounter{page}{1}
\pagestyle{plain}

\begin{bibunit}


\begin{abstract}
This paper investigates Bayesian variable selection when there is a hierarchical dependence structure on the inclusion of predictors in the model.
In particular, we study the type of dependence found in polynomial response surfaces of orders two and higher, whose model spaces are required to satisfy weak or strong heredity conditions. These conditions  restrict the inclusion of higher-order terms depending upon the inclusion of lower-order parent terms. We develop classes of priors on the model space, investigate their theoretical and finite sample properties, and provide a Metropolis-Hastings algorithm for searching the space of models.  The tools proposed allow fast and thorough exploration of model spaces that account for hierarchical polynomial structure in the predictors and provide control of the inclusion of false positives in high posterior probability models.
\end{abstract}


\section{Background}

In modern regression problems, choosing subsets of good predictors from a large pool is standard practice --- especially considering the pervasiveness of big data problems. Such analyses require multiple testing corrections in order to attain reasonable Type I error rates as well as control for false positives. This multiple testing problem is further complicated when it is critical to incorporate interactions or powers of the predictors.  Not only can there be a large number of predictors, but the structure that underlies the space of predictors must be taken into account. For regressions with a large number of predictors or high degree, the complete model space is too large to enumerate, and automatic stochastic search algorithms are necessary to find appropriate parsimonious models.

More than two decades years ago, \citet{P_90} exposed the relevance of respecting polynomial hierarchy among covariates in the variable selection context. To make model selection invariant to coding transformations in polynomial interaction models (e.g., to centering of the main effects), the selection must be constrained to the subset of models that fully respect the polynomial hierarchy \citep{GRS_82,P_87,P_90,MN_89,N_00,K_02}; such models satisfy a \textit{strong heredity condition}  (SHC).  Succinctly, a model satisfies SHC if for any predictor in the model every lower-order predictor associated with it is also in the model; for example, the model using the intercept, $\bx_1$, $\bx_2$ and $\bx_1\bx_2$. Such models are referred to as \textit{strong heredity models} (SHMs).

For the inclusion of a given term, a relaxation of the SHC requires that only a subset of lower-order terms is also included. This broader class of models does not attain the invariance properties of SHMs but can represent scientifically valid assumptions \citep{N_98}. A particular class of this type of model satisfies a \text{weak heredity condition} (WHC) and models in this class are referred to as \textit{weak heredity models} (WHMs). SHMs are examples of WHMs, but the WHM class is larger. A model including the intercept, $\bx_1$, and $\bx_1\bx_2$ is a WHM, but not a SHM. Specific definitions of the heredity conditions for defining SHMs and WHMs in the context of this paper are given in Section 2.

Although research on this topic started more than three decades ago \citep{Nelder77}, only recently have modern variable selection techniques been adapted to account for the constraints imposed by the polynomial hierarchy. As described in \citet{Bien2013}, the current literature on variable selection for polynomial response surface models can be classified into three broad groups: multistep procedures \citep{BSC_09,P_87}, regularized regression methods \citep{Yuan2009,Choi2010,Bien2013}, and Bayesian approaches \citep{Ch_96,Chipman1997,Chipman1998}.  This paper addresses the analysis of models satisfying a heredity condition (HC) from a Bayesian perspective. 

Contributing to the Bayesian literature on model selection, we focus on both the effect of the assumption of a HC as well as the influence of the prior for the model space. We prove a new theoretical result that provides a further argument in favor of the SHC over the WHC. We show that under the SHC assumption for a model space the posterior asymptotically concentrates on a single model, whereas this need not be the case for the WHC. We then modify and extend the model space priors developed in \citet{Ch_96} to provide classes of prior distributions incorporating assumptions about independence or exchangeability of term inclusion. 

The Bayesian variable selection problem consists of comparing models $\M$ in a model space $\cM$ using their posterior probabilities, given by $p(\M|\by,\cM)\propto m(\by|\M)\pi(\M|\cM)$.
Model posterior probabilities depend on the model space prior as well as on the priors for the model-specific parameters, implicitly through the marginals $m(\by|\M)$. Priors on the model-specific parameters have been extensively discussed in the literature \citep{J_61,Zellner1980,Kass1996,BP_96, George2000, Berger2001, LPMCB_08}. In contrast, the effect of the prior on the model space has until recently been neglected.  \citet{Scott2010} highlighted the impact of the priors on the model space in the context of multiple testing. The Ockham's-razor effect implicit in Bayesian variable selection through the Bayes factor does not correct for multiple testing. This penalization acts against more complex models but does not account for the cardinality or structure of the model space. The required multiplicity penalty is``hidden away'' in the model prior probabilities $\pi(M|\mathcal{M})$ \citep{Scott2010}. \citet{wilson2010bayesian} motivated the need for stronger penalization than pure combinatorics in the context of big data.  Adequately formulating priors on the model space can both account for structure in the predictors and provide additional control on the detection of false positive terms. In contast, using the popular uniform prior over the model space may lead to the undesirable and ``informative'' implication of favoring the set of models models that include half of the number of terms in the largest model in $\cM$ \citep{wilson2010bayesian}.

When conducting variable selection within these model spaces, two relevant issues must be taken into account.  First,  the notion of model complexity takes on a new dimension. Complexity is not exclusively a function of the number of predictors, but also depends upon the depth and connectedness of the predictor associations defined by the polynomial hierarchy. Second, because the model space is defined by such relationships, stochastic search algorithms used to explore the models must also conform to these restrictions.  

To the best of our knowledge, 
this is the only Bayesian framework for exploring polynomial model spaces using priors that incorporate penalties for both the number of predictors and for the complexity introduced by the polynomial structure of the model space. Although we focus on polynomial response surface regression models with independent homoscedastic normal errors, the proposed methods can be extended to generalized linear models, linear mixed models, and generalized linear mixed models. The only difficulty in such extensions is the computation of Bayes factors, which are most easily achieved for the homoscedastic normal regression problem. Instead, we focus on the prior structures that can be defined on the model space itself and the effect of these priors on subsequent inference.

 The manuscript is organized as follows. In Section 2, we review the hierarchical structure of polynomial regression and associated model spaces in terms of directed acyclic graphs (DAGs), facilitating the definition of the priors and random walks on the model spaces.  We propose, investigate, and provide recommendations for types of prior distributions on the model spaces in Section 3.  Section 4 provides a Metropolis-Hastings algorithm that takes advantage of the heredity structure.  In Section 5, the performance of the proposed methods is explored through simulation studies. We illustrate the algorithm on real data and compare the results with other procedures from current literature in Section 6.  Finally, Section 7 contains recommendations and discussion.  Supplementary materials contain details about the methods used and additional simulation results.

\section{Polynomial models and model spaces} 
 
 Suppose that the observations $\by=(y_1,\ldots,y_n)$ are modeled using polynomial regression
 \begin{equation} \label{eq:model}
\by=\sum \beta_{(\alpha_1,\ldots,\alpha_\p)} \prod_{j=1}^\p \bx_{j}^{\alpha_j} +\boldsymbol{\epsilon},
 \end{equation}
where $(\alpha_1,\ldots,\alpha_\p)=\balpha\in\N^\p$, $\N$
is the set of natural numbers including $0$,  $\mbs{\epsilon}\sim N(0,\sigma^2\bI)$, and only a finite number of nonzero regression coefficients. As an illustration, consider a model that includes  polynomial terms incorporating covariates $\bx_{1}$ and $\bx_{2}$ only.  The terms $\bx_{2}^2$ and $\bx_{1}^2\bx_{2}$ can be represented by $\balpha=(0,2)$ and $\balpha=(2,1)$, respectively.

The set $\N^\p$ constitutes a partially ordered set. It is ordered through the binary relation ``$\preceq$'' between pairs $(\balpha,\balpha')$, which is defined by $\balpha^\prime\preceq\balpha$ whenever $\alpha_j\geq\alpha_j^\prime$ for all $j=1,\ldots,\p$. Additionally, $\balpha^\prime\prec\balpha$ if $\balpha^\prime\preceq\balpha$ and $\alpha_j>\alpha^\prime_j$ for some $j$.   The order of a term $\balpha\in\N^p$ is given by the sum of its elements, $\order(\balpha)=\sum \alpha_j$, the {\it{length}} of $\balpha$ is given by the number of nonzero $\alpha_j$, and the {\it{type}} of $\balpha$ is defined by the increasing sequence of nonzero elements of $\balpha$.  If $\order(\balpha)=\order(\balpha')+1$ and $\balpha^\prime\prec\balpha$ then $\balpha^\prime$ is said to immediately precede $\balpha$, which is denoted by $\balpha^\prime\rightarrow\balpha$.  The {\it parent} set of $\balpha$ is defined by $\parents(\balpha)=\{\balpha^\prime\in\N^\p\ :\ \balpha^\prime\rightarrow\balpha\}$, which is the set of nodes that immediately precede $\balpha$. Similarly, the children set of $\balpha$ is defined as 
$\children(\balpha)=\{\balpha^\prime\in\N^\p\ :\ \balpha\rightarrow\balpha^\prime\}$. The poset $\N^\p$ can be represented by a directed acyclic graph (DAG), denoted by $\graph(\N^\p)$, with directed edges to a node from its parents.

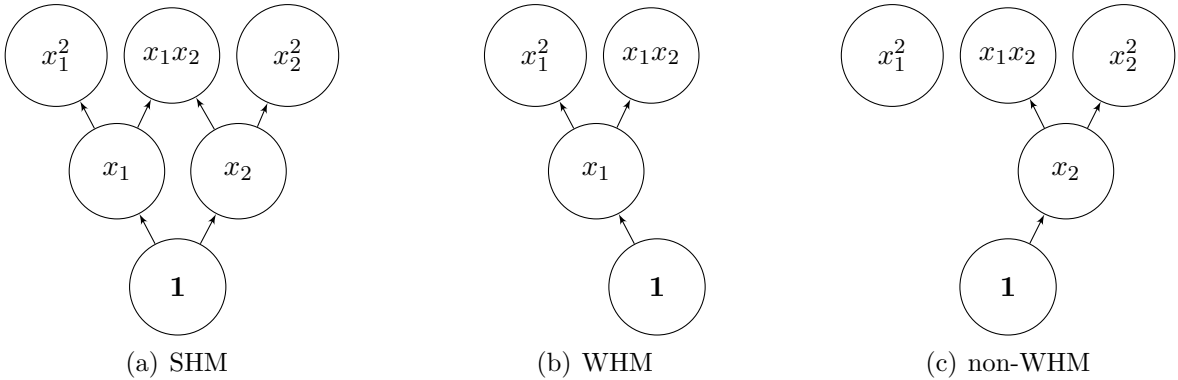
\begin{figure}[h]
\centering
\small{
\subfigure[SHM]{
\begin{tikzpicture}
\node[block3](Int){$\1$};
\node[block3,above of=Int,xshift=-2.1em](x1){$x_{1}$};
\node[block3,above of=Int,xshift=2.1em](x2){$x_{2}$};
\node[block3,above of=x1,xshift=-2.1em](x1sq){$x_{1}^{2}$};
\node[block3,right of=x1sq](x1x2){$x_{1}x_{2}$};
\node[block3,right of=x1x2](x2sq){$x_{2}^{2}$};
\path[line](Int)--(x1);
\path[line](Int)--(x2);
\path[line](x1)--(x1sq);
\path[line](x1)--(x1x2);
\path[line](x2)--(x1x2);
\path[line](x2)--(x2sq);
\end{tikzpicture}
 \label{fig:WFMsp}
}\qquad\qquad\subfigure[WHM]{
\begin{tikzpicture}
\node[block3](Int){$\1$};
\node[block3,above of=Int,xshift=-2.1em](x1){$x_{1}$};
\node[block3,above of=x1,xshift=-2.1em](x1sq){$x_{1}^{2}$};
\node[block3,right of=x1sq](x1x2){$x_{1}x_{2}$};
\path[line](Int)--(x1);
\path[line](x1)--(x1sq);
\path[line](x1)--(x1x2);
\end{tikzpicture}
 \label{fig:WFMsp}
}\qquad\qquad\subfigure[non-WHM]{   
\begin{tikzpicture}
\node[block3](Int){$\1$};
\node[block3,above of=Int,xshift=2em](x2){$x_{2}$};
\node[block3,above of=x2,xshift=-2em](x1x2){$x_1x_{2}$};
\node[block3,right of=x1x2](x2sq){$x_{2}^{2}$};
\node[block3,left of=x1x2](x1sq){$x_{1}^{2}$};
\path[line](Int)--(x2);
\path[line](x2)--(x1x2);
\path[line](x2)--(x2sq);
\end{tikzpicture}
 \label{fig:ChSet}
 }}
 \caption{Examples of graphs for different heredity conditions.}
 \label{fig:wfm}
\end{figure}

A polynomial model $\M$ satisfies the SHC if $\balpha\in \M$ implies that $\parents(\balpha)\subset\M$, and $\M$ satisfies the WHC if $\balpha\in\M$ implies $\balpha^\prime\in\M$ for some $\balpha^\prime\in\parents(\balpha)$.   Any SHM $\M$ is represented by a subgraph $\graph(\M)$ of $\graph(\N^\p)$ with the property that if node $\balpha\in\graph(M)$, then all nodes pointing to $\balpha$ are also in $\graph(\M)$.  A WHM graph can include a node $\balpha$ if there is a path from the intercept node to $\balpha$. Figure \ref{fig:wfm} illustrates instances of these models.

\setcounter{subfigure}{0}

The motivation for considering only SHMs is compelling.  The subspace of $\by$ modeled by a polynomial surface, given by the hat matrix $\bH_\M$, is invariant to linear transformations of the covariates used in the surface if and only if $\M$ corresponds to a SHM \citep{P_90}.  For example, if $\p=2$ and $\by=\beta_{(0,0)}+\beta_{(1,0)}\bx_{1}+\beta_{(0,1)}\bx_{2}+\beta_{(1,1)}\bx_{1}\bx_{2}+\bepsilon$, then the hat matrix is invariant to any covariate transformation of the form $({\bx_{1}\ \bx_{2}})\mbf{A}+\1\mbf{b}^\prime$ for any diagonal real nonsingular $2\times2$ matrix $\mbf{A}$ and any real column vector $\mbf{b}$ of length two.  In contrast, if $\by=\beta_{00}+\beta_{20}\bx_{1}^2+\bepsilon$,  the hat matrix formed after applying the transformation $\bx_{1}\mapsto \bx_{1}+\1c$ for real $c\neq0$ is not the same as the hat matrix formed by the original $\bx_{1}$.  This will make the selection depend on how the variables are coded. \citet{N_98} demonstrated that the conditions, under which weak heredity allows the design matrix to be invariant to coding, are seldom met exactly in practice. A WHM that does not obey the SHC constrains the regression model by enforcing ``special points'' (e.g. intercept fixed to zero or zero gradient point specified a priori). Special points arise in situations where omitting lower-order terms is justified due to the nature of the data and are valid in restricted scientific contexts, but they are the exception rather than the rule \citep{N_98}. Theorems \ref{thm:SHC} and \ref{thm:WHC} in Section 2.1 provide further theoretical justification for restricting attention to SHMs.
 
The spaces of models, $\cM$, considered in this paper are characterized by two SHMs: $\MB$, the base model, and $\MF$, the full model, which we assume is fixed and finite.	The base model $\MB$ consists of terms that are not subject to selection and is nested in the full model $\MF$. For example, $\MB$ can incorporate covariates describing the structure of an experimental design.  The model space $\cM$ is populated by all models $\M$ that satisfy the desired HC, contain $\MB$, and are nested in $\MF$.  To define the models explicitly, let $\kset(\M)=\M\setminus \MB$ and define the sets of {\it extreme} nodes and {\it children} nodes of $\M$ by $\extreme(\M)=\{\balpha\in\kset(M)\ :\ \M\setminus\{\balpha\}\text{ satisfies the HC}\}$ and $\children(\M)=\{\balpha\in\kset(\MF)\setminus\kset(M)\ :\ \M\cup\{\balpha\}\text{ satisfies the HC}\}$, respectively.
Under the SHC, $\M\in\cM$ is uniquely determined by considering either $\extreme(\M)$ or $\children(\M)$. Under the WHC, $\M\in\cM$ is determined through knowing both $\extreme(\M)\text{ and }\children(\M)$.

\subsection{Theoretical posterior considerations}
We now provide results regarding the posterior of the model space when the true model $\MT$ is nested in $\MF$ and we assume a HC for $\cM$; the results are all asymptotic as $dim(\by)=n\rightarrow\infty$. We require that the design matrix of the full model, defined as $\bX_{Full}$ with higher-order terms included, satisfies $\frac{1}{n}\bX_{Full}^\prime\bX_{Full}\rightarrow \Omega$, where $\Omega$ is positive definite. If a factor is considered as a covariate, the inclusion of one of its levels forces the inclusion of all of its levels; columns are dropped when collinearity due to factor variables or their interactions is present.

The Bayes factors for homoscedastic normal regression are obtained using intrinsic priors. The learning rate of model $\MT$ versus model $\M$ is exponential in $n$ when $\MT\nsubseteq\M$ and proportional to $n^{(|\M|-|\MT|)/2}$ when $\MT\subseteq\M$. Specifically, we have (up to $O(1)$)
\[
BF_{\M_T,\M}(\by)\rightarrow \left\{
\begin{array}{cc}
n^{\frac{|\M|-|\M_T|}{2}}\exp{\frac{n}{2}\log{\left(1+\Delta_{\M_T,\M}\right)}}& \text{ if }\M_T\nsubseteq\M\\
n^{\frac{|\M|-|\M_T|}{2}}\exp{-\frac{1}{2}\chi^2_{|\M|-|\M_T|}}& \text{ if }\M_T\subseteq\M,\\
\end{array}\right.
\]
where 
$
\Delta_{\M_T,\M}=\lim_{n\rightarrow\infty}\bbeta^\prime_{\M}\bX^\prime_{\M}(\bI-\bH_{\M_T})\bX_{\M}\bbeta_{\M}/(n\sigma^2_{\M})
$
is a directed distance of $\M_T$ from $\M$ \citep{Giron2010}. Note that these learning rates are achieved across large classes of parametric Bayesian models with fixed prior distributions, making the results of these theorems quite generalizable \citep{Kass1995}. In the theorems, we assume that the prior probabilities of all models satisfying the assumed HC are nonzero; the asymptotic results are independent of the particular positive prior assumed for $\cM$.

\begin{Theorem}
\label{thm:SHC}
Suppose that $\M_T\subseteq \MF$ is the true model, which may or may not be in $\cM$, and assume that $\cM$ satisfies the SHC. Let $\M_{\tilde{T}}=\{\balpha^\prime\in\MF\ :\ \balpha^\prime\preceq\balpha \text{ for some }\balpha\in\M_T\}\cup \MB$. Then $\MT\subseteq\M_{\tilde{T}}\in\cM$ and $p(\M_{\tilde{T}}|\by,\bX,\cM)\rightarrow 1$.
\end{Theorem}

The result in Theorem \ref{thm:SHC} depends only on the fact that the SHC is assumed for $\cM$ and not on whether $\M_T$ satisfies a HC (only that $\MT\subseteq\MF$). Note that if $\MT\in\cM$ then $\M_{\tilde{T}}=\MT$.  When only the WHC is assumed for $\cM$, a similar result holds.

\begin{Theorem}
\label{thm:WHC}
Suppose that $\M_T\subseteq \MF$ is the true model, which may or may not be in $\cM$, and assume that $\cM$ satisfies the WHC. Let $t=\min\{|\M_{\tilde{T}}|\ :\ \M_T\subseteq \M_{\tilde{T}}\in\cM\}$ and $\cM_T=\{\M_{\tilde{T}}\ :\ \M_T\subseteq \M_{\tilde{T}}\in\cM\text{ and }|\M_{\tilde{T}}|=t\}$. Then $p(\cM_T|\by,\bX,\cM)\rightarrow 1$. Further, if $\M_T$ satisfies the WHC, then $|\cM_T|=1$. 
\end{Theorem}

In Theorem \ref{thm:WHC}, it is interesting to note what happens when $\MF\supseteq\M_T\notin\cM$. If $\cM$ satisfies the WHC, there can be more than one model in $\cM_T$. The Bayes factor between models $\M,\M^\prime\in\cM_T$ is asymptotically given by $BF_{\M,\M^\prime}\rightarrow\exp{\frac{1}{2}\left(\chi^2_{\M^\prime,\M_T}-\chi^2_{\M,\M_T}\right)}$, where the $\chi^2$ distributions have $t-|\M_T|$ degrees of freedom and are not necessarily independent. Thus, though the posterior probability of $\cM_T$ does converge to $1$, no unique model is selected unless $|\cM_T|=1$. An example of this comes from a true model given by $\by=\beta_{(1,1)} \bx_1\bx_2+\bepsilon$. In this case, the two smallest models satisfying the WHC that nest $\M_T$ are given by $\M_{\tilde{T},1}=\{(0,0),(1,0),(1,1)\}$ and $\M_{\tilde{T},2}=\{(0,0),(0,1),(1,1)\}$. Because $\by$ is not correlated with $\bx_1$ or $\bx_2$ conditioned on the inclusion of $\bx_1\bx_2$, the asymptotic posterior probabilities of $\M_{\tilde{T},1}$ and $\M_{\tilde{T},2}$ are random.

These theorems amount to Cromwell's Rule and a statement about Bayesian learning rates. The posterior concentrates on a set of models $\cM_T$ which are of the smallest size, contain the true model $\M_T$, and have nonzero prior probability. The difference between Theorem \ref{thm:SHC} and Theorem \ref{thm:WHC} comes from the fact that the SHC places zero prior probability on models that satisfy the WHC but not the SHC. This provides another theoretical argument in favor of the SHC over the WHC. Under the SHC, the posterior concentrates on a single best model within the model space, whereas realizations of the $\chi^2$ random variables can produce arbitrary distinctions between the models in $\cM_T$ under the WHC.

\section{Priors on the model space}
In this section, we develop different prior structures on model spaces defined by a HC, discuss their advantages and disadvantages, and describe reasonable choices for their hyperparameters.   Our hyperparameter choices are motivated by the multiplicity prior from \citet{Scott2010} and the penalization prior of \citet{wilson2010bayesian}. We investigate how the choices of prior structure and hyperparameter affect the prior and posterior probabilities for predictor inclusion.

\subsection{Model prior definition}
For $\balpha\in\kset(\MF)$, let $\gamma_\balpha(\M)$ be the indicator function describing whether $\balpha$ is included in $\M$.  Let $\bgamma^\nu(\M)=\{\gamma_\balpha(M):\order(\balpha)=\nu\}$ and $\bgamma^{<\nu}(\M)=\bigcup_{j=0}^{\nu-1} \bgamma^j(\M)$.  With these definitions, the prior probability of any model $\M\in\cM$ can be factored as 
$
\pi(\M|\cM)=\prod_{j=J^{\min}_\cM}^{J^{\max}_\cM} \pi(\bgamma^j(\M)|\bgamma^{<j}(\M),\cM)\;,
$
\noindent where $J_\cM^{\min}$ and $J_\cM^{\max}$ are, respectively, the minimum and maximum orders of nodes in $\kset(\MF)$, and $\pi(\bgamma^{J^{\min}_\cM}(\M)|\bgamma^{<J^{\min}_\cM}(\M),\cM)= \pi(\bgamma^{J^{\min}_\cM}(\M)|\cM)$.
\citet{Ch_96} simplifies prior distributions on $\cM$ based on two assumptions.  First, if $\balpha\text{ and }\balpha^\prime$ are order $j$ then $\gamma_\balpha$ and $\gamma_{\balpha^\prime}$ are assumed to be {\it conditionally independent} given $\bgamma^{<j}$.  Second, the author invokes {\it immediate inheritance}: if $\order{(\balpha)}=j$ then $\pi(\gamma_\balpha(\M)|\bgamma^{<j}(\M), \cM)=\pi(\gamma_\balpha(\M)|\bgamma_{\parents(\balpha)}(\M), \cM)$, where $\bgamma_{\parents(\balpha)}(\M)$ is the inclusion indicator vector for $\parents(\balpha)$. 

Let $\pi_\balpha(\M)=\pi(\gamma_\balpha(\M)=1|\bgamma_{\parents(\balpha)}(\M),\cM)$ be the conditional inclusion probability of node $\balpha$ in model $\M$.  Under the assumptions of conditional independence and immediate inheritance, the prior probability of $\M$ is $\pi(\M|\bpi_\cM,\cM)=\prod \pi_\balpha(\M)^{\gamma_\balpha(\M)}(1-\pi_\balpha(\M))^{1-\gamma_\balpha(\M)}$,
with $\bpi_\cM=\lrb{\pi_\balpha(\M):\balpha\in\kset(\MF),\ \M\in\cM}$. Under the SHC, $\pi_\balpha(\M)=\gamma_\balpha(\M)=\mbf{0}$ if $\bgamma_{\parents(\balpha)}(\M)=0$, and under the WHC, $\pi_\balpha(\M)=\gamma_\balpha(\M)=0$ if $\prod_{\balpha^\prime\in\parents(\balpha)}\gamma_{\balpha^\prime}(\M)=0$.  Thus, the product can be restricted to the set of nodes $\balpha\in\kset(\M)\cup\children(\M)$.  Structure can be built into the prior on $\cM$ by making assumptions about the inclusion probabilities $\pi_\balpha(\M)$, such as equality assumptions or assumptions of a hyper-prior for these parameters. We now elaborate upon five model prior definitions, which incorporate various assumptions about the term inclusion indicators and the probabilities of term inclusion. Graphical representations of the priors are provided in Appendix A of the Supplementary materials.

\paragraph{Hierarchical Uniform Prior (HUP)}
The HUP assumes that the non-zero probabilities $\pi_\balpha(\M)$ are all equal.  Specifically, for a model $\M\in\cM$ it is assumed that $\pi_\balpha(\M)=\pi$ for all $\balpha\in\kset(\M)\cup\children(\M)$.  A full Bayesian specification of the HUP is completed by assuming a prior distribution for $\pi$.  The choice of $\pi\sim Beta(a,b)$ produces
$
\pi^{HUP}(\M|\cM,a,b)=B(|\kset(\M)|+a,|\children(\M)|+b)/ B(a,b)
$
where $B$ is the beta function.  
The HUP assigns equal probabilities to all models for which the sets of nodes $\kset(\M)$ and $\children(\M)$ have the same cardinality. This prior yields combinatorial penalization, but essentially fails to account for the hierarchical structure of the model space, as models with the same number of terms and children get the same probability irrespective of depth and connectedness of the model's graph.  An additional penalization for model complexity can be incorporated into the HUP by changing the values of $a$ and $b$.  Because $\pi_\balpha(\M)=\pi$ for all $\gamma_\balpha(\M)$ that are not forced to be zero, this penalization can only depend on characteristics of the entire graph for $\MF$.  One such penalization is to take $\ba=1$ and $\bb=|\kset(\MF)|$ for every model in $\cM$.

\paragraph{Hierarchical Independence Prior (HIP)}
The HIP assumes that there are no equality constraints among the non-zero $\pi_\balpha(\M)$.  Each non-zero $\pi_\balpha(\M)$ is given its own prior, which we assume to be Beta($a_\balpha,b_\balpha$).  The prior probability of $\M$ under the HIP is
$
\pi^{HIP}(\M|\cM,\ba,\bb)=\left(\prod_{\balpha\in \kset(\M)} \frac{a_\balpha}{a_\balpha+b_\balpha}\right)\left(\prod_{\balpha\in \children(\M)} \frac{b_\balpha}{a_\balpha+b_\balpha}\right)
$
, where the product over the empty set  is taken to be one.  Because the $\pi_\balpha(\M)$ are independent, any choice of $a_\balpha$ and $b_\balpha$ is equivalent to choosing a probability of success $\pi_\balpha(\M)$ for a given $\balpha$.  Under the SHC, the HIP with parameters $a_\balpha=b_\balpha=1$ is equivalent to the prior proposed in \citet{Ch_96} with fixed conditional inclusion probability of $1/2$ for each term.
This choice of hyperparameters accounts for the hierarchical structure of the model space, but essentially provides no penalization for combinatorial complexity at different levels of the hierarchy.  This can be observed by considering this choice when all terms in $\MF$ are order one, which is the equal probability prior (EPP).

An additional penalization for model complexity can be incorporated into the HIP.  In the construction of the prior, each vector of inclusion indicators for terms of order $j$, $\bgamma^j(\M)$, is conditioned on the  vector of inclusion indicators for lower order terms, $\bgamma^{<j}(\M)$. Therefore, $a_\balpha$ and $b_\balpha$ for $\balpha$ of order $j$ can be specified as functions of $\bgamma^{<j}(\M)$.  One such additional penalization utilizes the number of nodes of order $j$ that could be added without violating the assumed HC.  We refer to this particular type of penalty by $ch$, representing its dependence on (potential) children sets.  Choosing $a_\balpha=1$ and $b_\balpha(\M)=ch_j(\bgamma^{<j}(\M))$ is equivalent to choosing a probability of success $\pi_\balpha(\M)=[1+ch_j(\bgamma^{<j}(\M))]^{-1}$.

\paragraph{Hierarchical Order Prior (HOP)}
A compromise between complete equality and complete independence of the $\pi_\balpha(\M)$ is to assume equality between the non-zero $\pi_\balpha(\M)$ of a given order and independence across the different orders.  Define $\kset_j(\M)=\{\balpha\in\kset(\M):\order(\balpha)=j\}$ and $\children_j(\M)=\{\balpha\in\children(\M):\order(\balpha)=j\}$.  The HOP assumes that $\pi_\balpha(\M)=\pi^{(j)}(\M)$ for all $\balpha\in\kset_j(\M)\cup\children_j(\M)$.  Assuming that $\pi^{(j)}(\M)\sim Beta(a_j,b_j)$ provides a prior probability of
$
\pi^{HOP}(\M|\cM,\ba,\bb)=\prod_{j=J^{\min}_\cM}^{J^{\max}_\cM}
({B(|\kset_j(\M)|+a_j,|\children_j(\M)|+b_j)})/({B(a_j,b_j)})
$.
The specific choice of $a_j=b_j=1$ for all $j$ produces a hierarchical version of the \citet{Scott2010} multiplicity correction. An additional complexity penalization can be incorporated into the HOP in a similar fashion to the HIP.  Given $\bgamma^{<j}(\M)$, the number of order-$j$ nodes that could be added is $ch_j(\bgamma^{<j}(\M))=|\kset_j(\M)\cup\children_j(\M)|$.  Using $a_j=1$ and $b_j(\M)=ch_j(\bgamma^{<j}(\M))$ produces a hierarchical version of the penalization introduced in \citet{wilson2010bayesian}.

The fully specified HOP amounts to replacing the conditional independence and immediate inheritance conditions with an order-based exchangeability condition. In particular, the nodes that give rise to $ch_j(\M)$ are exchangeable for each order $j$, and the $\bgamma^j(\M)$ are independent across different orders (assuming that they satisfy the choice of HC for $\cM$).

\paragraph{Hierarchical Length Prior (HLP) and Hierarchical Type Prior (HTP)}
Incorporating additional exchangeability assumptions might be of interest to a researcher who, for example, would like to account for the fact that nodes with more connections in a model's graph should be penalized differently from nodes that have fewer connections.  To incorporate this type of additional structure we propose the HLP and the HTP. These two priors are equivalent if $J^{\max}_{\cM}\leq3$ and can differ if $J^{\max}_{\cM}\geq4$.

In particular, the HLP is constructed by making different group exchangeability conditions than the HOP. The $\bgamma^{j}(\M)$ are assumed independent, but the nodes in $\kset_j(\M)\cup\children_j(\M)$ are taken to be group exchangeable and not merely exchangeable. The sets of nodes in $\kset_j(\M)\cup\children_j(\M)$ are given by $(\kset_j(\M)\cup\children_j(\M))_{\ell}=\{\balpha\in \kset_j(\M)\cup\children_j(\M)\ :\ length(\balpha)=\ell\}$. The sets of nodes $(\kset_j(\M)\cup\children_j(\M))_{\ell}$ are independent across the different length groups but the nodes within a given $(\kset_j(\M)\cup\children_j(\M))_{\ell}$ are exchangeable. In other words, nodes of a given order with the same number of parents in $\MF$ are assumed to be exchangeable, conditioned on their inclusion satisfying the HC of $\cM$.

Similarly, the HTP is constructed by assuming node groups defined by $(\kset_j(\M)\cup\children_j(\M))_{t}=\{\balpha\in \kset_j(\M)\cup\children_j(\M)\ :\ type(\balpha)=t\}$. Thus, nodes of a given order of the same type are assumed to be exchangeable. Consider a full order four polynomial surface in two variables. The nodes $\bx_1^3\bx_2$ and $\bx_1\bx_2^3$ are grouped together, the nodes $\bx_1^4$ and $\bx_2^4$ are grouped together, and $\bx_1^2\bx_2^2$ is grouped by itself. Both the HLP and HTP can incorporate complexity penalizations in the fashion of the HIP or HOP, with $ch$ representing the number of nodes available for a given group that is assumed to be exchangeable.

\paragraph{Further penalizations in WHMs}
In his discussion of WHMs, \citet{Ch_96} specifies conditional probabilities on the inclusion of a given node by taking into account the number of parents of that node that are included in the model. As an example, consider the inclusion of $\bx_1\bx_2\doteq(1,1)$ in a model $\M$. The probability $\pi_{(1,1)}(\M)$ can take four values based upon the inclusion indicators for its parent terms $\bx_1$ and $\bx_2$. Call these probabilities $\pi_{(1,1)}(0,0)$, $\pi_{(1,1)}(1,0)$, $\pi_{(1,1)}(0,1)$, and $\pi_{(1,1)}(1,1)$. Under the WHC, the author sets the probabilities to be $0, 0.25, 0.25, 0.5$, reflecting the belief that the more parents of $\bx_1\bx_2$ included in $\M$, the higher $\pi_{(1,1)}(\M)$ ought to be. Such a belief can be incorporated into the HIP by letting $a_\balpha$ and $b_\balpha$ depend on the number of included and missing parents of $\balpha$. For example, letting $a_\balpha=1$ and $b_\balpha=1+2[|\parents(\balpha)|-\sum_{\balpha^\prime\in\parents(\balpha)}\gamma_{\balpha^\prime}(\M)]$ produces the four probabilities that the author set for $\pi_{(1,1)}(\M)$ assuming the WHC.

Similarly, a penalization based on the number of included and missing parents can be incorporated into the HLP and HTP. These priors already group nodes based upon their order and the size of their parent sets. The nodes can be further grouped by the number of parents that are missing from $\M$ and $a_\balpha$ and $b_\balpha$ can be modified depending on $\bgamma_{\parents(\balpha)}(\M)$.

\subsection{Prior sensitivity to distributional choices}

Each form of the priors introduced in Section 3.1 defines a whole family of model priors, characterized by the probability distribution for the vector of inclusion probabilities $\bpi_\cM$.  Here we compare the two choices of hyperparameters described in Section 3.1 for a specific model space. The first assumes that the parameters of the beta distributions are all $(1,1)$ and it referred to as $\ba=\1,\bb=\1$.   The second alternative is referred to as $\ba=\1,\bb=ch$, where $\bb=ch$ has different interpretations depending on the prior and represents a (potential) children based penalty. Table \ref{fig:priorvals1} shows the values in the prior for an order-two surface in two predictors when the base model $\MB$ is taken to be the intercept-only model and we have assumed the SHC a for $\cM$. Though this example is certainly not exhaustive, it does help build some intuition for how the different priors and hyperparameter choices behave.

\begin{table}[h!]
\begin{center}
\begin{tabular}{cc|cc|cc|cc|cc}\toprule
&\multirow{2}{*}{Model}&\multicolumn{2}{c}{HIP}&\multicolumn{2}{|c|}{HOP}&\multicolumn{2}{c}{HUP}
&\multicolumn{2}{c}{HLP/HTP}\\
 && $(\1,\1)$ & $(\1,ch)$ & $(\1,\1)$ & $(\1,ch)$ & $(\1,\1)$ & $(\1,ch)$ & $(\1,\1)$ & $(\1,ch)$\\ 
\cline{1-10}
1&1&$1/4$&$4/9$&$1/3$&$1/2$&$1/3$&$5/7$& $1/3$ &$1/2$\\
2&$1,\bx_1$&$1/8$&$1/9$&$1/12$&$1/12$&$1/12$&$5/56$& $1/12$ &$1/12$\\
3&$1,\bx_2$&$1/8$&$1/9$&$1/12$&$1/12$&$1/12$&$5/56$ & $1/12$ &$1/12$\\
4&$1,\bx_1, \bx_1^2$&$1/8$&$1/9$&$1/12$&$1/12$&$1/12$&$5/168$& $1/12$ &$1/12$\\
5&$1,\bx_2, \bx_2^2$&$1/8$&$1/9$&$1/12$&$1/12$&$1/12$&$5/168$ & $1/12$ &$1/12$\\
6&$1,\bx_1, \bx_2$&$1/32$&$3/64$&$1/12$&$1/12$&$1/60$&$1/72$& $1/18$ &$1/24$\\
7&$1,\bx_1, \bx_2, \bx_1^2$&$1/32$&$1/64$&$1/36$&$1/60$&$1/60$&$1/168$& $1/36$ &$1/72$\\
8&$1,\bx_1, \bx_2, \bx_1\bx_2$&$1/32$&$1/64$&$1/36$&$1/60$&$1/60$&$1/168$& $1/18$ &$1/24$\\
9&$1,\bx_1, \bx_2, \bx_2^2$&$1/32$&$1/64$&$1/36$&$1/60$&$1/60$&$1/168$& $1/36$ &$1/72$\\
10&$1,\bx_1, \bx_2, \bx_1^2, \bx_1\bx_2$&$1/32$&$1/192$&$1/36$&$1/120$&$1/30$&$1/252$& $1/36$ &$1/72$\\
11&$1,\bx_1, \bx_2, \bx_1^2, \bx_2^2$&$1/32$&$1/192$&$1/36$&$1/120$&$1/30$&$1/252$& $1/18$ &$1/72$\\
12&$1,\bx_1, \bx_2, \bx_1\bx_2, \bx_2^2$&$1/32$&$1/192$&$1/36$&$1/120$&$1/30$&$1/252$& $1/36$ &$1/72$\\
13&$1,\bx_1, \bx_2, \bx_1^2, \bx_1\bx_2, \bx_2^2$&$1/32$&$1/576$&$1/12$&$1/120$&$1/6$&$1/252$& $1/18$ &$1/72$\\
\bottomrule
\end{tabular}
\end{center}
\caption{Prior probabilities for the space of SHMs associated with the quadratic surface in two main effects.}
\label{fig:priorvals1}
\end{table}

First, compare the priors for $(\ba,\bb)=(\1,\1)$.  The HIP induces a complexity penalization that only accounts for the order of the terms in the model. Models including $\bx_1$ and $\bx_2$ (models $6$ through $13$) are given the same prior probability and no penalization is incurred for the inclusion of any of the quadratic terms.  The HUP induces a penalization for model complexity that does not adequately penalize models for including additional terms; $\MF$ is given more probability than any model containing at least one term.  This lack of penalization for the full model is a consequence of it being the only model in $\cM$ of that particular size. That is, this model space distribution favors the base and full models as both are the only models in their respective size classes.  Similar behavior is observed with the HOP.  As models become larger, they are penalized for the increased number of terms.  However, once the models become large enough, the number of models of those particular sizes is reduced, producing less combinatorial penalization. In this example, the HLP and HTP coincide and produce more penalization for the inclusion of a square term than the interaction term.

In contrast, if $(\ba,\bb)=(\1,ch)$, all the priors produce strong penalization as models become more complex, both in terms of the number and order of the nodes contained in the model.    For all of the priors, adding a node $\balpha$ to a model $\M$ to form $\M^\prime$ produces $p(\M)\geq p(\M^\prime)$.  However, there are clear differences between the priors.  The HIP penalizes the full model the most, with the HLP/HTP penalizing it the least. This observation is not unique to this particular choice of $\MF$. Because the HLP/HTP divides the nodes of a given order into smaller groups than the HUP or HOP, the former will always produce less penalization under the $(\1,ch)$ hyperparameter choice than the latter.

\subsection{Posterior sensitivity to the choice of prior}
To explore posterior sensitivity to the choice of prior distribution, we performed a simulation experiment with a full order-two surface on five predictors. We assume the SHC for the model space with $\MB$ only including the intercept, which gives 38,619 models. The true model contains the three main effects, two square terms, and two interactions.
We simulated twenty datasets and consider results for model selection using EPP, HIP$(\ba,\bb)$, HUP$(\ba,\bb)$, and HOP$(\ba,\bb)$ (where $\ba=\bb=\mathbf{1}$ and $\ba=\mathbf{1}, \bb=\mathbf{ch}$). 

\begin{figure}[h!]
\centering
\includegraphics[height=10cm]{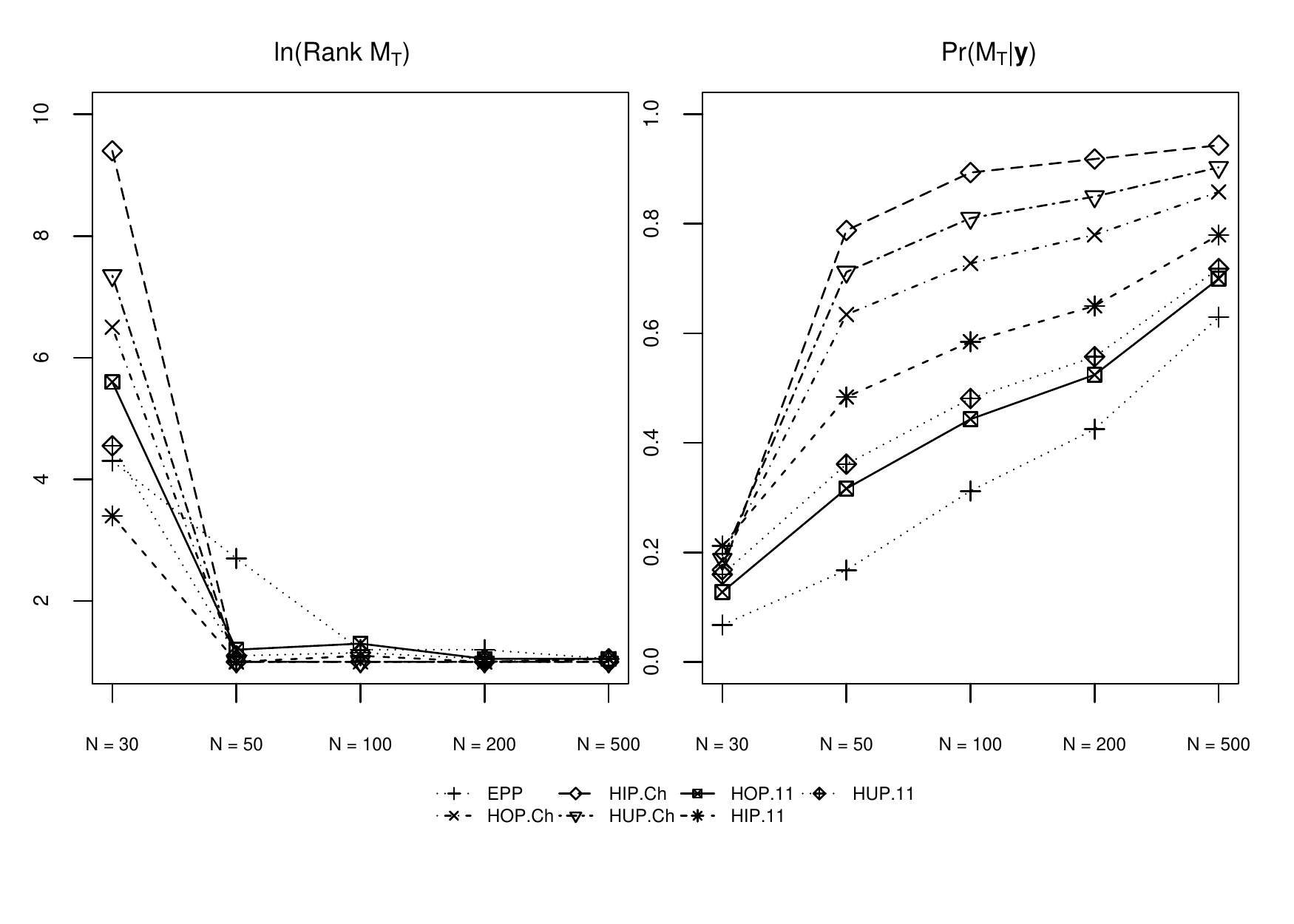}
\caption{Comparison of $\text{Rank }\MT$ and $p(\MT|\by)$ for different prior choices}\label{fig:mods33}
\end{figure}

Figure \ref{fig:mods33} shows the average rank of the true model and its average posterior probability over the twenty datasets for varying sample sizes. A general trend emerges that the EPP provides the least evidence in favor of the true model, while the choice of $\bb=ch$ provides the most evidence in favor of the true model when the sample size is moderate or large. However, when the sample size is small, the EPP and $\bb=1$ choices provide better rank for the true model. This is due to the additional penalization in the $\bb=ch$ priors, which needs to be overcome by the evidence contained in the Bayes factor.

\begin{table}[ht]
\centering
\begin{tabular}{lrrrrr l rrrrr}
  \toprule
\multirow{2}{*}{$n$}&\multicolumn{5}{c}{\%True Positives}& &\multicolumn{5}{c}{\%False Positives}\\
 & 30 & 50 & 100 & 200 & 500 & & 30 & 50 & 100 & 200 & 500 \\ 
\cline{1-6}\cline{8-12}
  SnS & 0.89 & 1.00 & 1.00 & 1.00 & 1.00 &   & 0.04 & 0.04 & 0.01 & 0.01 & 0.00 \\ 
  hierNet & 0.87 & 1.00 & 1.00 & 1.00 & 1.00 &   & 0.29 & 0.45 & 0.31 & 0.23 & 0.12 \\ 
  EPP & 0.91 & 1.00 & 1.00 & 1.00 & 1.00 &   & 0.03 & 0.07 & 0.02 & 0.03 & 0.00 \\ 
  HOP.Ch & 0.73 & 1.00 & 1.00 & 1.00 & 1.00 &   & 0.01 & 0.00 & 0.00 & 0.00 & 0.00 \\ 
  HIP.Ch & 0.68 & 1.00 & 1.00 & 1.00 & 1.00 &   & 0.01 & 0.00 & 0.00 & 0.00 & 0.00 \\ 
  HUP.Ch & 0.65 & 1.00 & 1.00 & 1.00 & 1.00 &   & 0.01 & 0.00 & 0.00 & 0.00 & 0.00 \\ 
  HOP.11 & 0.82 & 1.00 & 1.00 & 1.00 & 1.00 &   & 0.03 & 0.02 & 0.01 & 0.00 & 0.00 \\ 
  HIP.11 & 0.88 & 1.00 & 1.00 & 1.00 & 1.00 &   & 0.01 & 0.00 & 0.01 & 0.00 & 0.00 \\ 
  HUP.11 & 0.82 & 1.00 & 1.00 & 1.00 & 1.00 &   & 0.02 & 0.00 & 0.01 & 0.00 & 0.00 \\ 
   \bottomrule
\end{tabular}
\caption{Method comparison for the mean percentage of true and false 
             positive terms identified through selection} 
\label{tab:TPFP33}
\end{table}

These simulations were also carried out for the spike-and-slab (SnS) prior \citep{Mitchell1988, Ishwaran2005} as well as the lasso for hierarchical interactions (hierNet) \citep{Bien2013}. As seen in Table \ref{tab:TPFP33}, the highest posterior probability model (HPM) for all methods nests the true model when $n>50$. The heredity-based priors exhibit essentially no false positives when $n>100$, but false positives are observed with both hierNet and SnS as well as the EPP. For instance, when $n=100$, SnS, hierNet, and EPP have average false positive rates of  $0.0143$, $0.3071$, and $0.0167$, respectively. In contrast, the hierarchical priors with $\bb=ch$ have no false positives and $\bb=1$ has false positive rate $0.0083$. The additional penalizations built into the hierarchical priors strongly control the false positive rate in model selection. Of course, there is a trade-off and these priors can have difficulty finding true positives in situations where the signal-to-noise ratio is weak.

\section{Random Walks on the Model Space}
When the model space $\cM$ is too large to enumerate, a Metropolis-Hastings (MH) algorithm can be utilized to generate a dependent sample of models from the model space posterior.    The structure of $\cM$ both presents difficulties and provides clues on how to build algorithms to explore it.  Several kernels can be implemented in such an MH algorithm, three of which are outlined in this section and are based on local, global, and intermediate steps.  Combining the different strategies \citep{tierney1994markov}, e.g., by mixing Markov chain kernels, allows the MH algorithm to explore the model space thoroughly and quickly.  

For homoscedastic normal regression using intrinsic priors, the marginal probability of the observed $\by$ can be computed efficiently for any proposed model $\M$. Because of this, we bypass the intricacies of reversible jump and pseudo-prior methods \citep{Green1995,Carlin1995}. We note that when good approximations to the marginals are available, for example through the Laplace approximation \citep{Tierney1986,DiCiccio1997}, then the random walk algorithm can be implemented using them.

In this section, we succinctly describe the stochastic search algorithm proposed to explore $\cM$.  We detail three kernels designed with distinct scopes.  Although the default setting yields an automatic and sensible specification for the search algorithm, details related to tuning are discussed. Finally, we provide a brief discussion on the estimators used for $p(M|\by)$.

\subsection{Proposal kernels}

\paragraph{Global Jump}
In order to keep the chain from getting stuck in a local mode, the global jump kernel generates independently at random a model from the prior distribution.  The MH correction is the Bayes factor for the proposed model against the current model.

\paragraph{Local Jump}
Given a current model $\M$, the set $\extreme(\M)\cup\children(\M)$ contains the nodes whose inclusion can be changed while maintaining the assumed heredity condition of $\cM$.  The local jump kernel implements a stochastic forwards-backwards proposal kernel conditioned on $\M$.  For each $\balpha\in\extreme(\M)$ the model $\M_\balpha$ is defined to be the model $\M\setminus\{\balpha\}$ and for each $\balpha\in\children(\M)$ it is defined to be $\M\cup\{\balpha\}$.  The proposal kernel is supported only on the set of models $\cM_{\M}=\{\M\}\cup\{\M_\balpha:\balpha\in\extreme(\M)\cup\children(\M)\}$. Let $|\cM_{\M}|=N$ and $p(\M_\balpha|\by,\bX,\cM_{\M})$ be the renormalized posterior probability of $\M_\balpha$ restricted to $\cM_{\M}$. $\M_\balpha$ is proposed with probability $p(\M_\balpha|\by,\bX,\cM_{\M})/2+1/(2N)$. The proposal kernel is biased towards models with higher posterior probability, and each model has probability at least $1/(2N)$ of being proposed. Thus, even models with small posterior probability are proposed during iterations of the random walk and the probability of proposing any one model is bounded above by $0.5(1+1/N)$. 

When at model $\M$, this local jump kernel uses $m(\by|\M_\balpha)$ for all $\M_\balpha\in\cM_{\M}$ in order to compute the proposal probabilities. One can store $m(\by|\M_\balpha)$ for use in future proposal kernels as well as a renormalization estimator of the full posterior distribution, even if $\M_\balpha$ is never proposed or visited during the random walk. Thus, the balanced random walk helps to keep the sampler around models with high posterior probability while this transition kernel provides information about models that are close to those models.

\paragraph{Intermediate Jump}
The intermediate jump proposes a model by making proposals at each order.  The algorithm either increases the order from $J_\cM^{\min}$ to $J_\cM^{\max}$ or decreases it. Suppose that he chain is currently at $\M$. The algorithm creates intermediate models $\M^\prime_k$ for each $J_{\cM}^{\min}\leq k\leq J_{\cM}^{\max}$. Set $\cM_{\M^\prime}^{k}=\{\M^\prime\}\cup\{\M^\prime_{\balpha}\ :\ \balpha\in(\extreme(\M^\prime)\cup\children(\M^\prime))\cap\kset_{k}(\MF)\}$, which is the set of models that could be formed by adding or subtracting nodes of order $k$ to $\M^\prime$ or remaining at $\M^\prime$. When proposing an intermediate model from $\cM_{\M^\prime}^{k}$, the proposal density utilized is the same as that for the local jump restricted to the proposal set.

If increasing the order, set $\M^\prime_{J_{\cM}^{\min}-1}=\M$ and for $k=J_{\cM}^{\min},\ldots,J_{\cM}^{\max}$, propose models $\M^\prime_{k}$ from $\cM_{\M^\prime_{k-1}}^{k}$ sequentially with $k$ increasing. The final proposed model is $\M^\prime=\M^\prime_{J_{\cM}^{\max}}$. The decreasing-order kernel is defined analogously and is the reverse of the increasing-order kernel.  The MH correction is calculated by accounting for the posterior probabilities of the proposed model $\M^\prime$ and current model $\M$ as well as the proposal probabilities, which are computed using the proposal probabilities for each order.

For these intermediate steps, the use of the local jump proposal restricted to nodes of a given degree provides a means of proposing more models than the standard local jump (and thus storing more marginals of $\by$ conditioned on specific models). The final proposed model $\M^\prime$ can differ from $\M$ by as little as no nodes or as many as $J_{\cM}^{\max}-J_{\cM}^{\min}+1$ nodes (one for each order). This allows quicker exploration of the model space (in terms of number of iterations) while requiring more computation at each iteration.

\paragraph{Algorithm Tuning}
The sampling algorithm outlined in this paper in two ways. First, the weights in the convex combination of the restricted posterior and restricted uniform proposal in the local and intermediate jumps can be modified. When the weight on the restricted posterior is near one, the kernel is biased towards proposing models with high posterior probability. Conversely, when the weight is near zero, the kernel proposes models uniformly at random from the restricted space. The tradeoff is between proposing models with a high probability of being accepted versus increasing the probability of proposing low probability models and potentially moving to another part of the model space that might also contain high probability models. As a default, weights of $1/2$ are used.

Second, the relative frequency of utilizing the global, local, and intermediate jumps as proposal kernels can be modified. The global jump has relatively low probability of proposing good models, but can provide large moves across the model space. In contrast, the local jump has high probability of proposing models that will be accepted, but will move slowly around the model space. The intermediate jump allows for larger moves than the local jump, but at larger computational costs. The local jump should be used the most often with the global jump used the least to ensure that the algorithm does not get stuck in a local neighborhood of the model space. As a default, the kernels are used with frequency $1/2$, $2/5$, and $1/10$ for the local, intermediate, and global jumps, respectively.

\subsection{Posterior Estimators}
The MH sampler is uniformly ergodic because the model space is finite, and one can employ either renormalization or visit frequency to compute posterior quantities of interest, such as model posterior probabilities  \citep{garcia2013sampling}. When the model space is large, the number of iterations required to obtain good estimators of posterior probabilities with visit frequencies can be too large to be practically feasible. 

An alternative renormalization framework can be implemented that utilizes not only models visited during the random walk, but also any models used in proposal kernels during the random walk. Kernels that utilize aspects of  the model space posterior provide the necessary marginals, $m(\by|\M,\cM)$, to compute estimators that incorporate a larger portion of the model space than visit frequency estimators. Thus, the implemented renormalization estimators provide smaller total variation distance to the posterior than obtained using frequency estimators. Of course, this produces an efficiency trade-off. The transition kernels that propose a subset of models $\cM_{\M}\subset \cM$ when at a model $\M$ can require the computation (or lookup) of $m(\by|\M^\prime)$ for all $\M^\prime\in\cM_{\M}$, which has more cost than simple random proposals.

To compare the ability of the algorithm to estimate the model posterior probabilities through the proposed renormalization approach and frequency-based estimation, we explore the posterior probabilities obtained from the random walk performed for the example in Section 3.3. In particular, we follow the approach suggested in \citet{Chib1995}, where the entire model space is enumerated and the true model posterior probabilities calculated for each model.  The true posterior probabilities were then compared to the estimated ones.

Table \ref{tab:TVD} shows the the total variation distance (TVD) between the resulting estimators and the actual model posterior probabilities.
In the least favorable scenario ($n=30$ and $5,000$ iterations) renormalization and frequency estimators produced a mean TVDs of 0.0079 and 0.2625, respectively.  In the most favorable scenario ($n=200$ and $50,000$ iterations) the resulting TVDs were 0.0007 and 0.2149.  Despite the uniform ergodicity of the sampler, the frequency estimates converge slowly whereas renormalization provides a good approximation to the true posterior probabilities.

\begin{table}[ht]
\centering
\begin{tabular}{l|rrrrrrr}
  \toprule
\multirow{3}{*}{$n$}&\multicolumn{7}{c}{number of iterations}\\
& \multicolumn{3}{c}{Renormalization}&&\multicolumn{3}{c}{Frequency}\\
 & 5,000 & 10,000 & 50,000 && 5,000 & 10,000 & 50,000\\ 
\cline{1-4}\cline{6-8}
30 & 0.0079 & 0.0073 & 0.0062 && 0.2625 & 0.2564 & 0.2540 \\ 
100 & 0.0013 & 0.0012 & 0.0012  && 0.2470 & 0.2406 & 0.2325 \\ 
200 & 0.0010 & 0.0008 & 0.0007  && 0.2165 & 0.2157 & 0.2149 \\ 
  \bottomrule
\end{tabular}
\caption{Mean TVD between true model posterior probabilities and estimators.}\label{tab:TVD}
\end{table}

\section{Simulation Study}

To study the operating characteristics of the proposed priors, a simulation experiment was designed to determine the effect of the sample size, the signal-to-noise ratio, signal allocation, heredity assumption for true model, and model space heredity assumption. The model space is defined with $\MB$ being the intercept-only model and $\MF$ being the complete order-four polynomial surface in five main effects, which has with 126 nodes.  The cardinality of this model space  is $|\cM|>1.2\times10^{22}$, which makes enumeration of all models infeasible. The entries of the matrix of main effects are generated as independent standard normal variates.  The response vectors are drawn from the $n$-variate normal distribution as  $\by\sim N_n\left(\bX_{\M_T}\bbeta_{\M_T},\bI_n\right)$, where $\M_T$ is the true model. The highest order term in the true model has order three.

\begin{figure}[h] 
\begin{center}
\includegraphics[scale=.4]{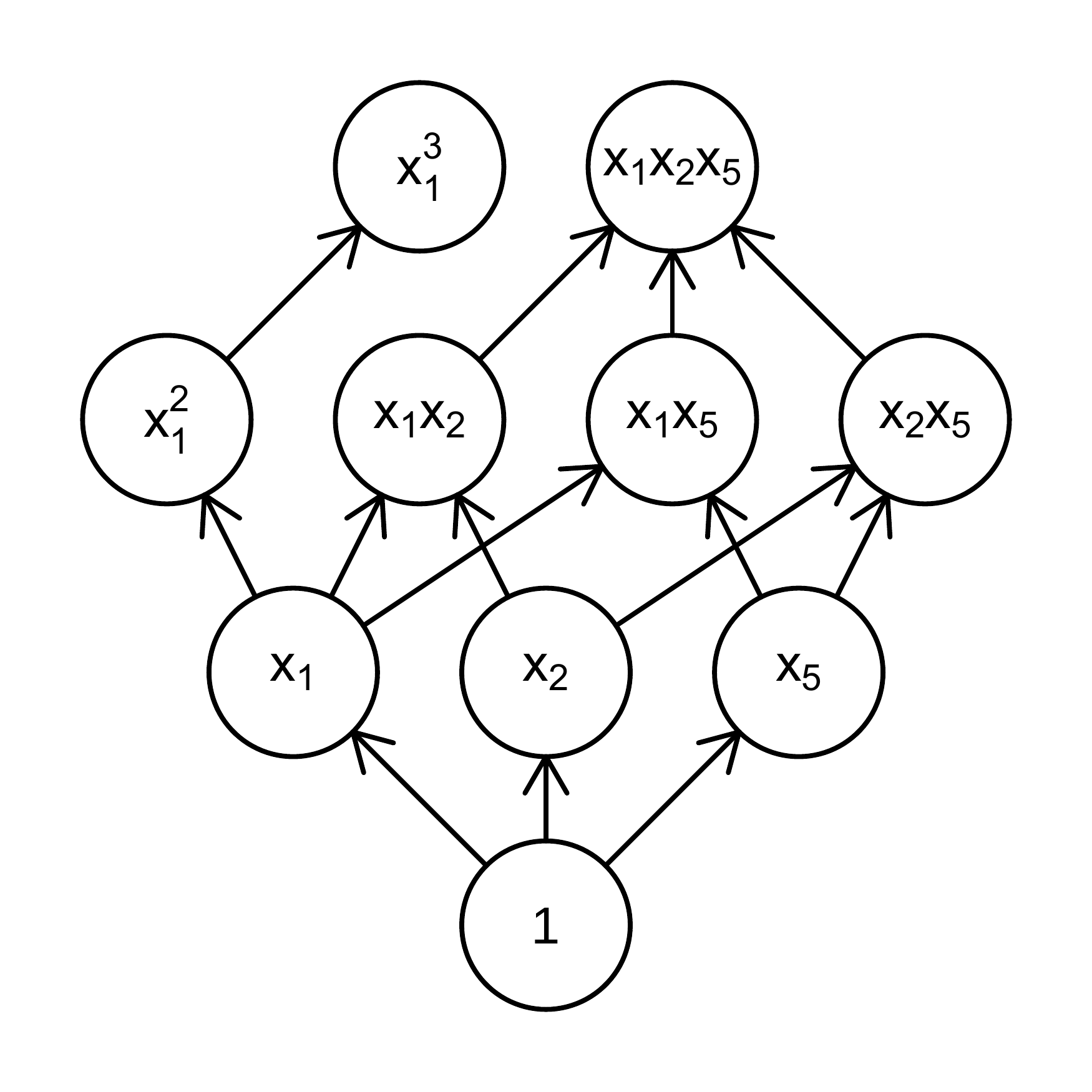}
\end{center}
\caption{$\M_T$: DAG of the largest true model used in simulations.}
\label{fig:TrueModSim}
\end{figure}

We consider three different values for $n$ ($130, 260, 1040$)  and signal-to-noise ratio (SNR= $0.25, 1, 4$). Three different signal allocations are investigated. The first places signal equally across orders one, two, and three. The second decreases the signal strength by half when increasing order. The third increases signal strength by a factor of two when increasing order. The largest true model considered is displayed in Figure \ref{fig:TrueModSim}, which satisfies the SHC. The second choice for $\M_T$ removes nodes $\bx_1\bx_2$ and $\bx_2\bx_5$ from the graph to form a WHM that is not a SHM. Similarly, the third choice of $\M_T$ removes $\bx_1^2$ and $\bx_2\bx_5$ from the graph, producing a model that does not satisfy the WHC. Regardless of the HC assumed for $\cM$, this choice of $\M_T$ leads to unique $\M_{\tilde{T}}\in\cM$ with smallest cardinality nesting $\MT$.

We drive a random walk through the model space using the HIP$(\1,\1)$ prior distribution and compute renormalization estimators of posterior probabilities for the HIP, HUP, HOP, HLP, and HTP priors with choice of $(\ba,\bb)\in\{(\1,\1),(\1,ch)\}$ as well as the EPP.
The total number of combinations considered for  SNR, sample size, regression coefficient values, and nodes in $\MT$ amounts to 162 different scenarios. Each scenario was run with 100 independently generated datasets  for $10,000$ iterations of the random walk.
The results presented in this section pertain to median results across the datasets regarding the true and false positive rates of the highest posterior probability model, posterior probability of the true model, and rank of the true model. Because small $n$ and/or low SNR lead to underfitting, we also consider the number of simulations in which the true model was not  visited. Figures providing the necessary simulation summaries are provided in Appendix D of Supplementary Materials.

\paragraph{SNR and sample size effect}
As expected, small sample sizes conditioned upon a small SNR impair the ability of the algorithm to detect true nonzero coefficients. Across the simulations with $n=130$ and SNR$=0.25$, only the EPP did a good job of recovering true positives. For example, assuming that $\MT$ is a SHM and that signal is allocated equally across orders, the EPP has median true positive rates of $2/3$ and $1/3$ when assuming the SHC and WHC for $\cM$, respectively. The false positive rates were $0.43\%$ and $0.36\%$ under the EPP, respectively. In contrast, the hierarchical priors produced no false positives, but only the HIP($\1,\1$) of \citet{Ch_96} combined with the SHC assumption for $\cM$ obtained a nonzero true positive rate, which was $1/9$.

For low SNR and large sample size ($n=1040$), all of the priors choose $\MT$ as the highest posterior model (HPM) under the SHC. The major differences come from the allocation of posterior probability to $\MT$. The hierarchical priors obtain the highest probability for $\MT$ with the HUP($\1,ch$), which is $87\%$. In contrast, the EPP only assigns $8\%$ posterior probability to $\MT$. Under the WHC, the true positive rate falls to a high of $78\%$, which is achieved under the EPP and the HTP($\1,ch$). The contrast comes from the ranks for $\MT$, which are $1,426$ and $582$, respectively. The best rank was $262$, achieved by HOP($\1,ch$). In the best case scenario ($n=1040$ and SNR$=4$), the true positive rate for the WHC did not increase in this scenario. Under these same conditions, the SHC provides perfect selection. These results together provide a strong argument for the SHC over the WHC. Under the SHC, the HIP($\1,\1$) performed the best when the SNR was low, followed by the HLP and HTP with the same hyperparameter choice for moderate sample sizes. The true model was not found $30\%$ of the time under the SHC and never found under the WHC.

\paragraph{Coefficient magnitude}
For this comparison, we focus on $n=260$ and SNR $=1$, but with $\MT$ satisfying the WHC and not the SHC. First, we consider the different priors under the SHC. When the signal is concentrated more on the higher-order terms or distributed evenly across the orders, the hierarchical priors perform admirably with the exception of HUP($\1,ch$). In general, ranking and posterior probability of the true model is improved slightly by setting $\bb=ch$, but the strong control that this choice provides over false positives decreased the median true positive rate for that prior. When the signal is concentrated towards lower-order terms, the HTP($\1,\1$) performed the best, but only achieved a $2/3$ true positive rate. Similar to the discussion for SNR and sample size, the WHC gives generally weaker inference than the SHC. The trends tend to follow those for the SHC assumption.
One notable difference is that the $\bb=ch$ priors perform better or comparable to their $\bb=\1$ counterparts, with the exception of the HOP.

A general pattern is seen with the hierarchical priors under the SHC or WHC: evidence in favor of the null is lowest when the signal is concentrated on lower-order terms. Intuitively, this empirical result makes sense. Because we are penalizing the addition of higher-order terms, their signal needs to be larger in order to increase the evidence in favor of inclusion from Bayes factors. Concentration of the signal on higher-order terms provides the most inferential power under the SHC due to its strict parental inclusion requirements.

\paragraph{Special points on the scale}
Across the range of examples, assuming the SHC provides better inference than assuming the WHC in terms of true positives, false positives, and posterior probability and rank of the true model. Even when the true model does not satisfy the SHC (or any HC at all), the HIP, HTP, and HLP under the SHC assumption for $\cM$ provide the most balanced inference in favor of the pseudo-true model $\M_{\tilde{T}}$. Subsequent inference about the influence of terms in $\M_{\tilde{T}}$ can be carried out through estimation and consideration of credible intervals. 

\section{Method comparison: Ozone data analysis}

We analyze the ozone data from \citet{BF_85} to assess the performance of our procedures. We follow the analysis of \citet{LPMCB_08}, though we only use intrinsic priors to obtain our Bayes factors. After removing observations with missing values, the data contains 330 observations. A subset of 165 observations was sampled uniformly at random and used for variable selection, and the remaining data were used for validation.  We predict daily measurements of maximum ozone concentration near Los Angeles with eight meteorological variables. The meteorological variables and their two-way interactions and squares are used in $\MF$, which has 44 terms. The base model, $\MB$, is the intercept-only model.  There are $7.1\times 10^{10}$ and $6.5 \times 10^{11}$ models under the SHC and WHC, respectively.

The priors are assessed through their predictive accuracy on the validation dataset with the prediction root mean squared error (PRMSE). The results are provided for both for the highest posterior probability model (HPM) and under model averaging of the 500 highest posterior probability models.  Additionally, we compare the performance of the priors developed in this paper to the lasso for hierarchical interactions \citep{Bien2013}, implemented using the R package {\tt hierNet}.  The penalty parameter was obtained by minimizing the estimated cross-validation error in a ten-fold cross-validation. Given that the penalty parameter chosen is influenced by the cross-validation sets chosen, the cross-validation procedure was replicated 100 times. The results of this analysis are contained in Table \ref{tab:BestMods}. The HPMs referenced in the Table \ref{tab:BestMods} and the marginal inclusion probabilities for all $44$ terms are given in Appendix E of the Supplementary materials.

\begin{table}[h]
\centering
\begin{tabular}{ccc ccc ccc}
  \toprule
  \multirow{2}{*}{Prior}&\multirow{2}{*}{pars}&\multirow{2}{*}{HC}& \multicolumn{2}{c}{PRMSE} & \multicolumn{2}{c}{posterior prob} & \multirow{2}{*}{$|M_{\text{HPM}}|$}& \multirow{2}{*}{$M_{\text{HPM}}$}\\
&  &  & HPM & Ave & $M_{\text{HPM}}$ & $\lrb{M_{(k)}}_{k=1}^{500}$ & &\\ 
  \midrule
\multirow{4}{*}{HIP} & \multirow{2}{*}{(1,ch)} & SH & 4.2 & 4.128 & 0.278 & 0.999 & 6 &$M_{(1)}$\\ 
 &  & WH & 4.593 & 4.483 & 0.712 & 0.999 & 3 &$M_{(2)}$ \\ 
 & \multirow{2}{*}{(1,1)} & SH & 4.2 & 4.129 & 0.278 & 0.998 & 6 &$M_{(1)}$ \\ 
 & & WH & 4.593 & 4.212 & 0.605 & 0.997 & 3 & $M_{(2)}$ \\ 
  \midrule
\multirow{4}{*}{HOP} & \multirow{2}{*}{(1,ch)} & SH & 4.2 & 4.141 & 0.249 & 0.991 & 6 & $M_{(1)}$ \\ 
 &  & WH & 4.593 & 4.125 & 0.412 & 0.985 & 3  & $M_{(2)}$ \\ 
 & \multirow{2}{*}{(1,1)} & SH & 4.2 & 4.141 & 0.251 & 0.993 & 6 & $M_{(1)}$ \\ 
 &  & WH & 4.593 & 4.125 & 0.412 & 0.986 & 3 &  $M_{(2)}$ \\ 
  \midrule
\multirow{4}{*}{HUP} & \multirow{2}{*}{(1,ch)} & SH & 4.2 & 4.12 & 0.256 & 0.993 & 6 & $M_{(1)}$\\ 
 &  & WH & 4.593 & 4.354 & 0.703 & 0.995 & 3 &  $M_{(2)}$ \\ 
 & \multirow{2}{*}{(1,1)} & SH & 4.2 & 4.121 & 0.256 & 0.994 & 6  & $M_{(1)}$ \\ 
 &  & WH & 4.593 & 4.195 & 0.635 & 0.99 & 3 &  $M_{(2)}$ \\ 
  \midrule
\multirow{4}{*}{HLP} & \multirow{2}{*}{(1,ch)} & SH & 4.2 & 4.142 & 0.237 & 0.991 & 6 & $M_{(1)}$ \\ 
 & & WH & 4.593 & 4.17 & 0.344 & 0.99 & 3 &  $M_{(2)}$ \\ 
 & \multirow{2}{*}{(1,1)} & SH & 4.2 & 4.143 & 0.237 & 0.991 & 6 & $M_{(1)}$ \\ 
 &  & WH & 6.459 & 4.223 & 0.267 & 0.98 & 4 & $M_{(3)}$ \\ 
  \midrule
\multirow{2}{*}{EPP} & \multirow{2}{*}{--} & SH & 4.2 & 4.151 & 0.024 & 0.423 & 6 & $M_{(1)}$ \\ 
 &  & WH & 4.21 & 4.161 & 0.018 & 0.359 & 5 & $M_{(4)}$ \\ 
  \midrule
 \multirow{2}{*}{hierNet} & \multirow{2}{*}{--} & SH & 4.349 & -- & -- & -- & 24 & $M_{(5)}$ \\ 
  &  &WH & 4.371 & -- & -- & -- & 22 & $M_{(6)}$\\
  \bottomrule
\end{tabular}\caption{Predicted RMSE under the SHC and WHC for the highest probability model (HPM) and under model averaging (Ave).}\label{tab:BestMods}
\end{table}

The main results can be summarized as follows. First, the lowest PRMSE for a HPM comes from the SHC with hierarchical priors or the EPP. Second, the hierarchical priors provide stronger posterior concentration than the EPP. Third, HPMs under the SHC yield a lower PRMSE than those obtained under the WHC. Fourth, models found with hierNet are considerably larger than those resulting from the Bayesian procedures. However, hierNet provides slightly larger PRMSEs, showing that they overfit the data.
Fifth, model averaging PRMSEs slightly improve upon those from the HPM's. Sixth, smaller selected models are obtained under the WHC, which one expects given its relaxed inclusion condition. However, these selected models have greater predictive PRMSE, showing another possible argument in favor of the SHC over the WHC.

Interestingly, the HPM PRMSE under the WHC for the HLP$(1,1)$ is the largest observed.  However, the model averaging PRMSE indicates that the full Bayesian analysis using this prior still behaves well. The second highest probability model is $\M_{(2)}$, which has a comparable posterior probability (0.20) to that of the HPM (0.267).

\section{Discussion}
In this paper, we investigated prior structures for polynomial regression assuming the strong and weak heredity conditions and developed random walk samplers to draw from the model space. We extended the priors developed in \citet{Ch_96} by completing the Bayesian specification as well as developing priors that utilize exchangeability conditions. These hierarchical priors have similar posterior behavior, with the HOP, HLP, and HTP providing hierarchical versions of the multiplicity correction and complexity penalizing priors, which help immensely with model selection when the polynomial surface has a high degree. When the number of main effects is small or the signal-to-noise ratio is low, the HIP provides better model selection properties. However, it is ill-suited for regressions with a large number of predictors.

In addition to these results from the simulations, Theorems \ref{thm:SHC} and \ref{thm:WHC} provide a further theoretical argument in favor of strong --- over weak --- heredity. Through the analysis of the ozone data, we can also see that the WHC leads to higher prediction RMSE than the SHC. This is related to the theoretical result, which shows that the SHC concentrates posterior probability on a unique model, whereas the WHC might not.

The computational algorithm described in the paper takes advantage of the hierarchical nature of the model space and provides for renormalization estimators that incorporate a large portion of the model space. Through a simulation experiment, we have shown that these estimators have smaller total variation distance to the true posterior than visit frequency estimators. We provide the software necessary to carry out a Metropolis-Hastings random walk on the space of these model spaces through the \texttt{R} package \texttt{varSelectIP}.

The methods in this paper can be expanded to generalized linear mixed models and other modeling contexts. For example, for binary data the methods proposed by \citet{Albert1993} can be implemented using intrinsic priors on the parameter space \citep{Leon-Novelo2012} and with the proposed hierarchical priors on the model space.

The theoretical results of Section 2.1 can be easily extended to encompass these classes of models under suitable regularity conditions. Efficient computation or accurate approximation of the marginal probability of the data under competing models is required to employ the MH algorithm we have developed. However, because the focus of the paper is on the model space prior, the methods discussed in this paper can be incorporated into more complicated setting where more elaborate MCMC methods are necessary.


\putbib
\end{bibunit}

\end{document}